\documentclass[aip,apl,amsmath,amssymb, reprint, floatfix]{revtex4-1}

\usepackage{graphicx}
\usepackage{bm}

\begin{document}

\preprint{AIP/123-QED}

\title[Sample title]{Minimization of ion micromotion using ultracold atomic probes}

\author{A. H\"arter}
\author{A. Kr\"ukow}
\affiliation{Institut f\"ur Quantenmaterie and Center for Integrated Quantum Science and Technology IQ$^\text
  {ST}$, Universit\"at Ulm, 89069 Ulm, Germany}
\author{A. Brunner}
\altaffiliation{Current address: 3. Physikalisches Institut, Universit\"at Stuttgart, 70569 Stuttgart, Germany}
\affiliation{Institut f\"ur Quantenmaterie and Center for Integrated Quantum Science and Technology IQ$^\text
  {ST}$, Universit\"at Ulm, 89069 Ulm, Germany}
\author{J. Hecker Denschlag}
\affiliation{Institut f\"ur Quantenmaterie and Center for Integrated Quantum Science and Technology IQ$^\text
  {ST}$, Universit\"at Ulm, 89069 Ulm, Germany}

\begin{abstract}
We report on a new sensitive method to minimize excess micromotion of an ion in a Paul trap. The ion is placed in an ultracold cloud of neutral Rb atoms in
which ionic micromotion induces atomic losses and heating. Micromotion is minimized
by applying static electric offset fields such that both loss and heating are minimized. We achieve a compensation precision as high as the best
compensation methods to date. In contrast to these methods, our scheme is applicable even for ions that cannot be optically probed. Furthermore, our method avoids
the formation of temporary patch charges which are a main issue for the long-term stability of micromotion minimization.
\end{abstract}

\maketitle

In the last decades there has been a tremendous progress in experiments with ions confined in Paul traps. Single or well-defined numbers of ions
have been prepared and manipulated on the quantum level
$\,$\cite{Lei03,Bla08,Haf08,Bla12}. Laser cooling and manipulation
of these strongly isolated quantum objects can then be used for precision spectroscopy, quantum simulation and quantum computation. For such experiments, control
over the ionic excess micromotion is a pre-condition. Furthermore, a young line of research investigates cold collisions between trapped ions and ultracold
neutral atomic gases$\,$\cite{Vul2008,Zip10,Smi10,Hall2011,Hud2011,Ran2012}. Here, excess micromotion sets the dominant energy scale and it needs to be compensated with high precision to
reduce the atom-ion collision energies to the mK regime and beyond.

Micromotion is a driven oscillatory motion of the ions in the rf field of the Paul trap. Ideally, the ion is trapped at a
node of the rf field, where micromotion is minimal. However, possible stray electric fields displace the ion from this location and into
trap regions with increased rf fields. These in turn increase the micromotion amplitudes inducing the so-called excess micromotion. This micromotion
contribution thus needs to be minimized by compensating the stray electric fields.
To date, a number of methods have been devised to accurately compensate excess micromotion, all of which employ resonant scattering of light off the ion.
Such methods employ, e.g. motional side-band spectroscopy$\,$\cite{Ber98,Raa00,Chu12}, photon correlation measurements$\,$\cite{Ber98,Pyk12}, precise
position detection of the ion while changing the rf confinement$\,$\cite{Ber98}, or
detection of ionic motional excitation when resonantly modulating the ion trap potential$\,$\cite{Nar11,Tan12}. For the ionic species typically used
in these experiments, resonant light at wavelengths below 500$\,$nm is needed. This is known to produce unstable patch charges on dielectric surfaces making frequent
readjustments of the compensation voltages necessary. Also, the laser-based compensation methods fail when working with
``dark" ions, such as Rb$^+$, where optical transitions are not accessible.

Here, we present a novel compensation method where ion micromotion is probed by immersing a single ion into ultracold $^{87}$Rb atom clouds.
The ion collides with the atoms at typical rates of several kHz. Since the atomic temperatures $T_\text{at}$ are on the order of 1$\,\mu$K,
the dominant energy scale for these collisions is set by the ion micromotion. Through the collisions, energy is transferred from the ion to the
atoms$\,$\cite{Zip10,Smi10,Zip11}. If the transferred energy is larger than the atom trap depth, the colliding atom will be lost. Otherwise, the
atom remains trapped and eventually rethermalizes with the rest of the cloud
leading to an increase of the atomic temperature. After several seconds of immersion and typically many thousand collisions we detect both the number
of remaining atoms and the final atomic temperature by standard absorption imaging techniques.
In an iterative process, we minimize the loss and heating of the atom cloud by applying electric compensation fields, thus minimizing excess micromotion.

\begin{figure}
\includegraphics[width=8.0cm] {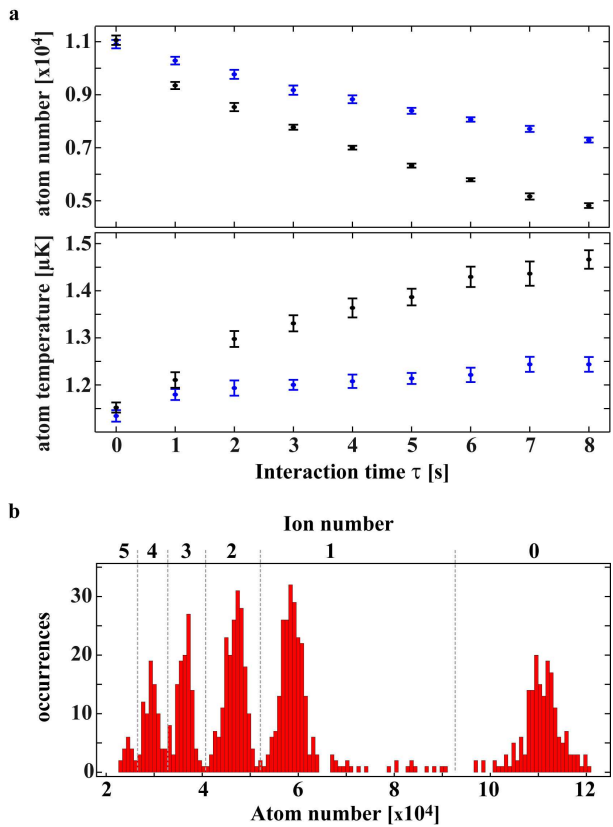}
\caption{\label{fig:1} Atomic signals after atom-ion interaction. (a) Evolution of
atom numbers (upper panel) and atomic temperatures (lower panel) during interaction
with a single ion. The measurement was performed both for $\varepsilon_r\lesssim 0.1\:$V/m
(blue data points) and for $\varepsilon_r= 4\:$V/m (black data points).
(b) Histogram of remaining atom numbers after exposing about 1000 atom clouds for 2$\,$s to a
variable number of trapped ions. The peaks in the distribution
correspond to the number of trapped ions. }
\end{figure}

The experiments are performed with $^{87}$Rb$^+$ ions confined in a linear Paul trap of which the design is discussed in detail in reference\cite{Smi12}.
The effective radial distance from the trap center to
the four rf electrodes is $R'=2.6\,$mm and the trap is driven at a frequency of $\Omega=2\pi\times4.17\,$MHz and an amplitude of $V_0=500\,$V. The endcap electrodes
are supplied with static voltages
of about 8$\,$V. This configuration results in trapping frequencies of $(\omega_r,\omega_z)= 2\pi\times(350,51)\,$kHz. To compensate radial ion micromotion,
we apply electric offset fields perpendicular to the trap axis by using two pairs of compensation electrodes.\\
We begin our investigations by immersing the ion into the center of a comparatively dilute atom cloud which is held in a far-detuned optical dipole trap \cite{Smi10}.
At atomic trap frequencies of $(136,141,40)\:\text{Hz}$,
initial atom number $N_\text{at}= 1.1\times10^4$ and temperature $T_\text{at}= 1.1\:\mu\text{K}$, the initial atomic peak density is
$n_\text{at}=1\times 10^{11}\:\text{cm}^{-3}$. The atom clouds are reproduced within an experimental cycle time of about 30$\,$s with fluctuations in atom number of less
than 5$\,$\%, even for thousands of experimental cycles.\\
Figure$\,$\ref{fig:1}a shows both decay and heating of the atom cloud as a function of the
interaction time $\tau$ when exposed to a single ion. This measurement was performed for two micromotion settings as determined by the radial electric offset field
$\varepsilon_r$. For a small offset field ($\varepsilon_r\lesssim 0.1\,$V/m, blue data points), atom loss and heating are suppressed as compared to a field of
$\varepsilon_r= 4\,$V/m (black data points). In addition to the electric fields, also the number of trapped ions obviously strongly affects the losses and
heating of the atom cloud. As we want to carry out all our minimization experiments with a single ion, we have developed a way to determine the ion
number in the cloud by simply looking at atomic losses.
Fig.$\,$\ref{fig:1}b shows a histogram of about 1000 atom loss experiments where atomic
clouds ($N_\text{at}= 1.1\times10^5$, $n_\text{at}= 3\times 10^{12}\:\text{cm}^{-3}$) were exposed to a variable number of trapped Rb$^{+}$ ions for 2$\,$s at
electric fields of several V/m. The well separated peaks of the distribution of the histogram reflect the number of trapped ions, as indicated in the graph.
Specifically, an atom number around $6\times10^4$ indicates the trapping of a single ion
which we then use for the further investigations of ion micromotion.

\begin{figure}
\includegraphics[width=8.0cm] {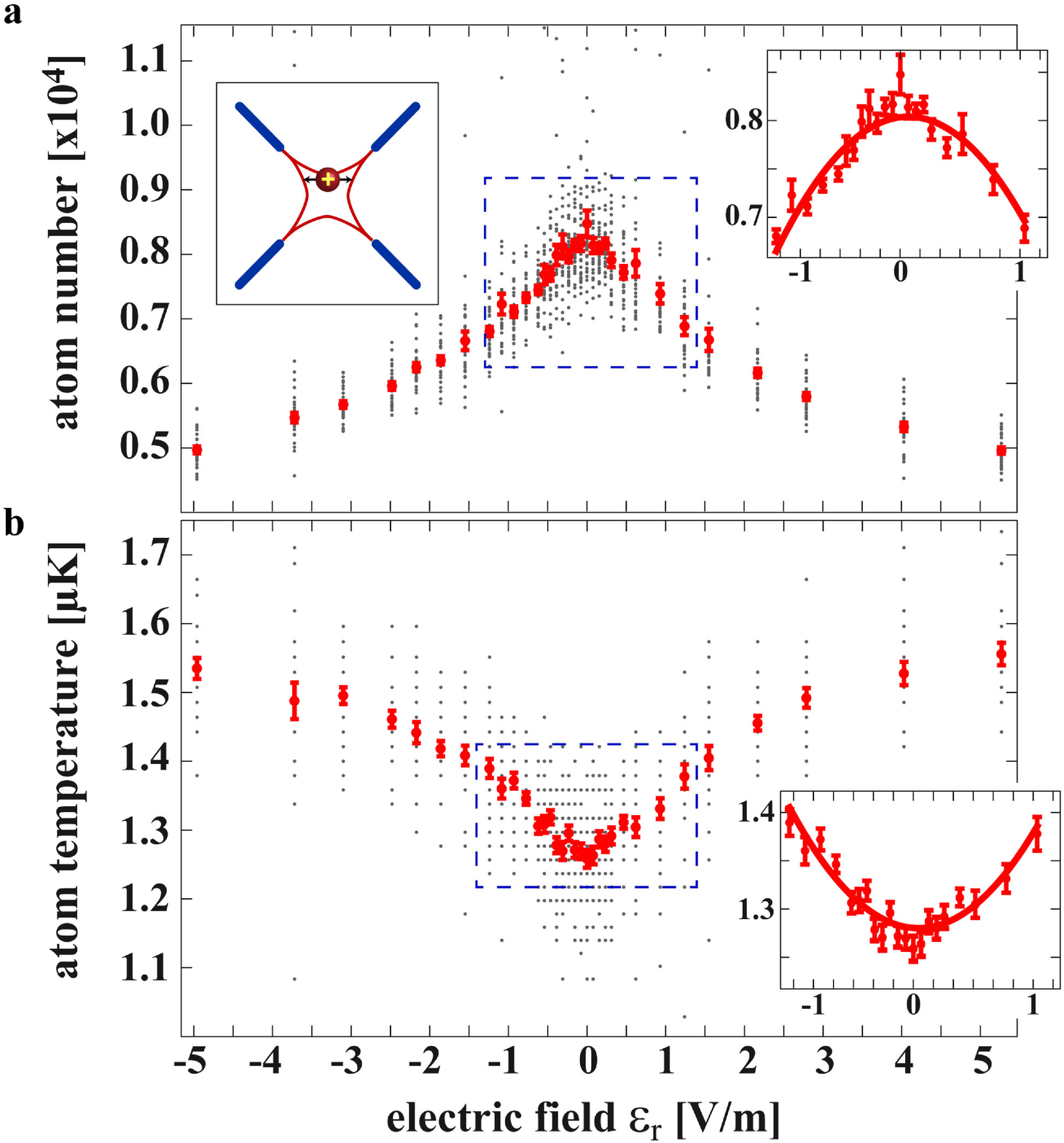}
\caption{\label{fig:2} Micromotion compensation using a dilute atom cloud.
The displayed mean atom numbers and temperatures include almost exclusively two-body atom-ion interactions.
\textit{Insets:} Parabolas are fitted locally to the atom number maximum and the
temperature minimum, respectively. This allows for electric field compensation down to
$\Delta\varepsilon_r\lesssim0.05\:$V/m. }
\end{figure}

We now perform a first micromotion compensation measurement for which we immerse the single ion for a fixed interaction time $\tau=8\,$s into the
dilute atom cloud ($N_\text{at}= 1.1\times10^4$, $n_\text{at}=1\times 10^{11}\:\text{cm}^{-3}$). We vary the radial electric fields between
$\varepsilon_r = \pm 5\:$V/m  and measure the final atom numbers and temperatures (Fig.$\,$\ref{fig:2}). At
each electric field setting we perform about 15 individual measurements which are shown as the scattered grey points in the figure. Each data point was measured with a freshly prepared atom cloud.
While most of these points lie within a relatively small range of scatter, there are some extreme outliers with almost no atom loss or heating
effect (best seen in Fig.$\,$\ref{fig:2}a). These extreme
events occur predominantly when ion micromotion is well compensated and are explained by three-body recombination processes between the ion and two neutral
atoms, as described in detail in $\,$\cite{Har12}. Briefly, the energy released upon recombination ejects the ion from the atom cloud
so that the atom-ion interaction temporarily stops. The ionic impact on the atom cloud is thus strongly reduced. In the measurement
shown in Fig.$\,$\ref{fig:2}, the atomic density is so low that such three-body interactions are very rare.

At this point of our investigations, we want to suppress their influence on the data. To do this, we first calculate the mean atom number and temperature
at each interaction time from all data points at this field setting. Then, we ignore those data points which lie outside a $2\sigma$ environment around these
mean values and average over the remaining measurement outcomes.
In this way, the influence of the extreme events is filtered out to a large part and the mean values given in Fig.$\,$\ref{fig:2} contain almost
exclusively two-body atom-ion collisions.\\
Figure$\,$\ref{fig:2} shows a monotonic dependence on
the electric field strength $|\varepsilon_r|$, both in the atom
numbers and the atomic temperatures. Atom losses and heating are
minimal for a vanishing offset field. Furthermore, as one might
expect, the data is symmetric with respect to the sign of the
electric field. Scans of this type can be used to minimize
electric fields with a sensitivity given by how precisely we can
determine the center of the peak (dip). Fitting a parabola to the data
(see insets in Fig.$\,$\ref{fig:2}) allows us to extract the optimal electric
field setting to within $\Delta\varepsilon_r\lesssim0.05\:$V/m.
The parameter of crucial importance in cold atom-ion interactions
is the corresponding micromotion energy$\,$\cite{Ber98}
\begin{equation}
E_{r}=\frac{m_\text{ion}}{16}(q_r r\, \Omega)^2\,,
\label{Eemm}
\end{equation}
where $m_\text{ion}$ is the ionic mass, $q_r=2eV_0/(m_\text{ion}
R'^2 \Omega^2)\approx0.24$ is the Mathieu parameter and
$r=e\varepsilon_r/(m_\text{ion} \omega_r^2)$ is the displacement
of the ion from the rf node. Using equation$\,$\ref{Eemm} we
derive a residual radial micromotion energy of $E_{r}=
k_\text{B} \times 3.2\mu$K from our uncertainty in $\varepsilon_r$.

\begin{figure}
\includegraphics[width=8.0cm] {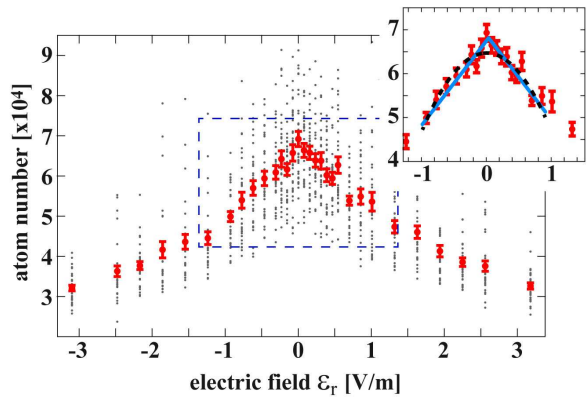}
\caption{\label{fig:3} Same as Fig.$\,$\ref{fig:2} but using an
atomic sample at a density of $n_\text{at}\approx1.5\times
10^{12}\:\text{cm}^{-3}$. The additional effects of three-body
atom-ion interaction increase the electric field sensitivity and
allow for field compensation down to
$\Delta\varepsilon_r\lesssim0.02\:$V/m. 
}
\end{figure}

It turns out that we can increase the sensitivity by carrying out
the measurement using a larger and denser atom cloud ($N_\text{at}
= 9\times 10^4$, $n_\text{at} = 1.5\times
10^{12}\:\text{cm}^{-3}$). The increase in density by more than an
order of magnitude compared to the previous measurement strongly
increases the three-body atom-ion recombination
rate$\,$\cite{Har12}. This can be seen when comparing the data scatter
in Figs.$\,$\ref{fig:3} and $\,$\ref{fig:2}a. Especially for
small electric fields $\varepsilon_r$ the scatter in Fig.$\,$\ref{fig:3}
is large and indicating that nearly every atom-ion interaction period
includes at least one three-body event. Thus, it does not make
sense to sort out data with three-body events. We simply take the
mean of all data points. Oddly, this changes the form of the data
curve (as compared to Fig.$\,$\ref{fig:2}a) which is now
cusp-shaped. The cusp can be explained by two facts:
1) Three body-recombination events in general lead to an
increase in the final atom number, as an ejected ion does not kick out atoms.
2) The probability for
three-body events increases strongly with decreasing micromotion.
As seen in the inset of Fig.$\,$\ref{fig:3}a, a parabola is not the
ideal fit to the cusp. A function such as $c_1
|\varepsilon_r - c_2| + c_3$, where $c_{1,2,3}$ are fit parameters,
does much better. We obtain an uncertainty of
$\Delta\varepsilon_r\lesssim0.02\:$V/m which corresponds to
micromotion energy $E_{r}= k_\text{B} \times 0.6\:\mu$K.

\begin{figure}
\includegraphics[width=8.0cm] {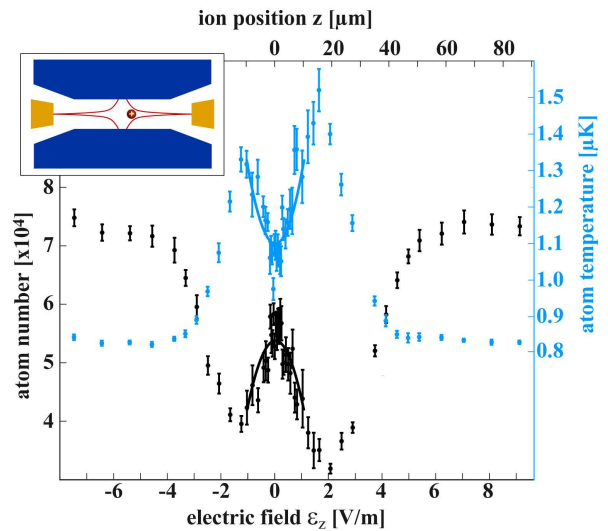}
\caption{\label{fig:4} Measurement of axial ion micromotion. Due to the small
value of $q_z$, the ion shifts through the atom cloud as its
micromotion is varied. In the outer part of the plot ($|z| \gtrsim 20\:\mu$m),
the density profile of the atomic sample
can be recognized. At smaller ionic displacements,
a drastic decrease of the impact on the atomic sample is observed.
The fits shown allow us to determine the position of the
rf frequency node to within $\Delta z\approx 0.6\:\mu$m.}
\end{figure}

The data shown in Figs.$\,$\ref{fig:1}-\ref{fig:3} are obtained by
varying electric offset fields in the vertical direction,
perpendicular to the trap axis. Measurements in the horizontal
direction are performed in a similar way with similar results. In
a symmetrically driven linear Paul trap, ideally, there is no
micromotion along the trap axis. We drive our trap in an
asymmetric way where two of the four rf electrodes are grounded. This
leads to non-vanishing rf fields (and micromotion) everywhere
along the trap axis but in the central point (see schematic in
Fig.$\,$\ref{fig:4}), analogous to the radial directions.
Although these axial rf fields are significantly weaker than
the corresponding radial ones, they still result in large micromotion
energies if the ion is strongly shifted from the trap center. Due to
the small axial trapping frequency of about $50\,$kHz, ions in our trap
are highly susceptible to electric stray
fields in axial direction. We adjust the axial offset
electric fields by changing the voltage on one of the
endcap electrodes. Figure$\,$\ref{fig:4} shows
a scan over several V/m which moves the ion through the entire atom cloud.
The inner parabolically shaped
parts of the data around $\varepsilon_r = 0$ are similar to the
curves in Figs.$\,$\ref{fig:2} and \ref{fig:3}. The outer wings,
however, mainly reflect the decrease of the density at the edge of
the atom cloud. Here, the ion probes the Gaussian density
distribution of the atoms (as discussed in ref.$\,$\cite{Smi10}). Again, fitting a
parabola to the central region of Fig.$\,$\ref{fig:4} we obtain
an electric field uncertainty of $\Delta \varepsilon_z\lesssim
0.06\,$V/m which corresponds to a positional accuracy of $\Delta
z\lesssim 0.6\,\mu$m and a residual axial micromotion energy
$E_\text{z}= k_\text{B}\cdot 9\,\mu$K$\;$\footnote{To calculate
$E_\text{z}$ we use equation$\,$\ref{Eemm} and replace
$q_r$ and $r$ by $q_z$ and $z$. In our trap $q_z\approx0.04\cdot
q_r\approx0.01$.}. Thus, the micromotion energy is significantly
larger than for the radial directions. Stronger axial
confinement of the ion would increase the achievable positional
accuracy of this measurement and thereby reduce the
corresponding micromotion energy.
 \\
\begin{table*}[t!]
\centering
\footnotesize\rm
\begin{tabular}{c| c | c | c | c | c | c | c | c | c | c}
Ion species & ref. & $\Delta\varepsilon_r$ [V/m] &  $u_r$ [nm] & $E_r$ [$\mu$K$\,k_\text{B}$] &
$\Delta\varepsilon_z$ [V/m] &  $u_z$ [nm] & $E_z$ [$\mu$K$\,k_\text{B}$] & $\Omega/2\pi\,$[MHz] & $\omega_r/2\pi\,$[MHz] & $\omega_z/2\pi\,$[MHz]\\
\hline
$^{138}$Ba$^+$ & \cite{Chu12} & 1.8 & 7 & 225 & - & - & - & 5.3 & 1.2 & 0.4\\
$^{172}$Yb$^+$ & \cite{Pyk12} & 0.9 & 1.1 & 60 & 0.3 & - & -  & 25.7 & 0.48 & 0.12 \\
$^{40}$Ca$^+$ & \cite{Nar11} & 0.4/2.5 & 6/40 & 380/17000 & - & - & -  & 15 & 1.2/1.4 & 0.4 \\
$^{27}$Al$^+$ & \cite{Cho10} & 1.1 & 0.4 & 18 & - & - & -  & 59 & 6 & 3\\
$^{88}$Sr$^+$ & \cite{Ake12} &  1.0 & 0.85 & 36  & - & - & -  & 22 & 2.2 & 1\\
$^{40}$Ca$^+$ & \cite{Chw09} & 0.42 & 0.46 & 3.6 & - & - & -  & 23.5 & 3.4 & 1.2 \\
$^{25}$Mg$^+$ & \cite{Hem11} & - & - & - & - & 0.2 & 0.75  & 25 & 4.5 & 2 \\
$^{87}$Rb$^+$ & & 0.02 & 0.5 & 0.5 & 0.06 & 24 & 9 & 4.17 & 0.35 & 0.05
\label{table:compvalues}
\end{tabular}
\caption{Comparison of commonly used figures of merit to quantify
ion micromotion. Residual electric fields $\varepsilon_{r,z}$,
micromotion amplitudes $u_{r,z}$ and micromotion
energies $E_{r,z}$ are given along with
the trap drive frequency and the secular frequencies. The first
seven rows show values extracted from information of the given
references. The last row gives the values obtained in this work.}
\end{table*}
The micromotion minimization scheme using atomic probes works in
all three spatial dimensions. This is a great advantage as it
relaxes the experimental complexity, e.g. in terms of optical
access to the trap center. Practically, however, a good
compensation in one direction requires an iterative process of
compensating all three dimensions. Only then the micromotion
energy of the ion can be significantly minimized. Indeed, the
data shown in this work were acquired after micromotion in the
remaining two dimensions had already been minimized.
\\
To verify that the atom-based micromotion compensation yields
the same optimal electric field settings as established
optical methods, we perform measurements on a single
$^{138}$Ba$^+$ ion with no atom cloud being present. We laser cool the ion and detect its
fluorescence on a charge-coupled device camera. We use two micromotion minimization methods:
1) minimizing position changes of the ion when changing the trap frequency of the
Paul trap \cite{Ber98}. 2) minimizing motional excitation of the ion while
modulating the rf with the trap frequency \cite{Nar11, Tan12}.
With these methods we are able to compensate radial electric fields to about
0.1$\:$V/m and to position the ion axially to better than 1$\:\mu$m (corresponding to
$\Delta \varepsilon_z\approx0.09\,$V/m).
\\
In order to benchmark our minimization method, we compare its accuracy to the
ones of various optical methods, as reported in the literature.
Table$\,$I lists the field uncertainties $\Delta \varepsilon_{r,z} $, the motional
micromotion amplitudes $u_{r,z}$, and the micromotion kinetic energies $E_{r,z}$ for a variety of species and ion traps.
Care has to be taken when directly comparing the results, as the set-ups, trap frequencies and rf-drive
frequencies differ substantially. Nevertheless, the table shows that atom-based micromotion compensation
is as good or even better than the best reported values achieved with
all standard methods. \\
In conclusion, we have presented a novel method to compensate ion excess micromotion in a Paul trap. The trapped ion is immersed into an atomic cloud.
Micromotion is detected in terms of atomic loss and heating of the cloud
as induced by two-body and three-body collisions between ion and atoms.
Our minimization results are as good as the best reported ones for fluorescence-based detection methods.
The method works in all three dimensions while requiring optical access from only a single direction. Besides compensation of excess micromotion due to
stray electric fields, as demonstrated here, it should also work against excess micromotion due to phase differences of opposing rf
electrodes of the Paul trap. The method can be used for all ionic species, including ``dark" ions without accessible optical transitions,
as long as collisions with the atomic gas are mainly elastic. Finally, as it does not involve optical ion detection, creation of patch
charges is avoided. Hence, long-term stability of the compensation settings is achieved, which is a prerequisite for precision compensation of micromotion.

The authors would like to thank Stefan Schmid and Wolfgang Schnitzler
for help in the lab and Michael Drewsen, Piet Schmidt, Ferdinand Schmidt-Kaler, Kilian Singer, David Hume for valuable information and fruitful discussions.
This work was supported by the German Research Foundation DFG within the SFB/TRR21.

\bibliography{micromin}

\providecommand{\noopsort}[1]{}\providecommand{\singleletter}[1]{#1}%
\begin{thebibliography}{24}%
\makeatletter
\providecommand \@ifxundefined [1]{%
 \@ifx{#1\undefined}
}%
\providecommand \@ifnum [1]{%
 \ifnum #1\expandafter \@firstoftwo
 \else \expandafter \@secondoftwo
 \fi
}%
\providecommand \@ifx [1]{%
 \ifx #1\expandafter \@firstoftwo
 \else \expandafter \@secondoftwo
 \fi
}%
\providecommand \natexlab [1]{#1}%
\providecommand \enquote  [1]{``#1''}%
\providecommand \bibnamefont  [1]{#1}%
\providecommand \bibfnamefont [1]{#1}%
\providecommand \citenamefont [1]{#1}%
\providecommand \href@noop [0]{\@secondoftwo}%
\providecommand \href [0]{\begingroup \@sanitize@url \@href}%
\providecommand \@href[1]{\@@startlink{#1}\@@href}%
\providecommand \@@href[1]{\endgroup#1\@@endlink}%
\providecommand \@sanitize@url [0]{\catcode `\\12\catcode `\$12\catcode
  `\&12\catcode `\#12\catcode `\^12\catcode `\_12\catcode `\%12\relax}%
\providecommand \@@startlink[1]{}%
\providecommand \@@endlink[0]{}%
\providecommand \url  [0]{\begingroup\@sanitize@url \@url }%
\providecommand \@url [1]{\endgroup\@href {#1}{\urlprefix }}%
\providecommand \urlprefix  [0]{URL }%
\providecommand \Eprint [0]{\href }%
\providecommand \doibase [0]{http://dx.doi.org/}%
\providecommand \selectlanguage [0]{\@gobble}%
\providecommand \bibinfo  [0]{\@secondoftwo}%
\providecommand \bibfield  [0]{\@secondoftwo}%
\providecommand \translation [1]{[#1]}%
\providecommand \BibitemOpen [0]{}%
\providecommand \bibitemStop [0]{}%
\providecommand \bibitemNoStop [0]{.\EOS\space}%
\providecommand \EOS [0]{\spacefactor3000\relax}%
\providecommand \BibitemShut  [1]{\csname bibitem#1\endcsname}%
\let\auto@bib@innerbib\@empty
\bibitem [{\citenamefont {Leibfried}\ \emph {et~al.}(2003)\citenamefont
  {Leibfried}, \citenamefont {Blatt}, \citenamefont {Monroe},\ and\
  \citenamefont {Wineland}}]{Lei03}%
  \BibitemOpen
  \bibfield  {author} {\bibinfo {author} {\bibfnamefont {D.}~\bibnamefont
  {Leibfried}}, \bibinfo {author} {\bibfnamefont {R.}~\bibnamefont {Blatt}},
  \bibinfo {author} {\bibfnamefont {C.}~\bibnamefont {Monroe}}, \ and\ \bibinfo
  {author} {\bibfnamefont {D.~J.}\ \bibnamefont {Wineland}},\ }\href@noop {}
  {\bibfield  {journal} {\bibinfo  {journal} {Rev. Mod. Phys.}\ }\textbf
  {\bibinfo {volume} {75}},\ \bibinfo {pages} {281} (\bibinfo {year}
  {2003})}\BibitemShut {NoStop}%
\bibitem [{\citenamefont {Blatt}\ and\ \citenamefont {Wineland}(2008)}]{Bla08}%
  \BibitemOpen
  \bibfield  {author} {\bibinfo {author} {\bibfnamefont {R.}~\bibnamefont
  {Blatt}}\ and\ \bibinfo {author} {\bibfnamefont {D.}~\bibnamefont
  {Wineland}},\ }\href@noop {} {\bibfield  {journal} {\bibinfo  {journal}
  {Nature}\ }\textbf {\bibinfo {volume} {453}},\ \bibinfo {pages} {1008}
  (\bibinfo {year} {2008})}\BibitemShut {NoStop}%
\bibitem [{\citenamefont {H\"affner}, \citenamefont {Roos},\ and\ \citenamefont
  {Blatt}(2008)}]{Haf08}%
  \BibitemOpen
  \bibfield  {author} {\bibinfo {author} {\bibfnamefont {H.}~\bibnamefont
  {H\"affner}}, \bibinfo {author} {\bibfnamefont {C.}~\bibnamefont {Roos}}, \
  and\ \bibinfo {author} {\bibfnamefont {R.}~\bibnamefont {Blatt}},\
  }\href@noop {} {\bibfield  {journal} {\bibinfo  {journal} {Phys. Rep.}\
  }\textbf {\bibinfo {volume} {469}},\ \bibinfo {pages} {155–203} (\bibinfo
  {year} {2008})}\BibitemShut {NoStop}%
\bibitem [{\citenamefont {Blatt}\ and\ \citenamefont {Roos}(2012)}]{Bla12}%
  \BibitemOpen
  \bibfield  {author} {\bibinfo {author} {\bibfnamefont {R.}~\bibnamefont
  {Blatt}}\ and\ \bibinfo {author} {\bibfnamefont {C.~F.}\ \bibnamefont
  {Roos}},\ }\href@noop {} {\bibfield  {journal} {\bibinfo  {journal} {Nature
  Physics}\ }\textbf {\bibinfo {volume} {8}},\ \bibinfo {pages} {277–284}
  (\bibinfo {year} {2012})}\BibitemShut {NoStop}%
\bibitem [{\citenamefont {Grier}\ \emph {et~al.}(2009)\citenamefont {Grier},
  \citenamefont {Cetina}, \citenamefont {Orucevic},\ and\ \citenamefont
  {Vuletic}}]{Vul2008}%
  \BibitemOpen
  \bibfield  {author} {\bibinfo {author} {\bibfnamefont {A.}~\bibnamefont
  {Grier}}, \bibinfo {author} {\bibfnamefont {M.}~\bibnamefont {Cetina}},
  \bibinfo {author} {\bibfnamefont {F.}~\bibnamefont {Orucevic}}, \ and\
  \bibinfo {author} {\bibfnamefont {V.}~\bibnamefont {Vuletic}},\ }\href@noop
  {} {\bibfield  {journal} {\bibinfo  {journal} {Phys. Rev. Lett.}\ }\textbf
  {\bibinfo {volume} {102}},\ \bibinfo {pages} {223201} (\bibinfo {year}
  {2009})}\BibitemShut {NoStop}%
\bibitem [{\citenamefont {Zipkes}\ \emph {et~al.}(2010)\citenamefont {Zipkes},
  \citenamefont {Palzer}, \citenamefont {Sias},\ and\ \citenamefont
  {K\"{o}hl}}]{Zip10}%
  \BibitemOpen
  \bibfield  {author} {\bibinfo {author} {\bibfnamefont {C.}~\bibnamefont
  {Zipkes}}, \bibinfo {author} {\bibfnamefont {S.}~\bibnamefont {Palzer}},
  \bibinfo {author} {\bibfnamefont {C.}~\bibnamefont {Sias}}, \ and\ \bibinfo
  {author} {\bibfnamefont {M.}~\bibnamefont {K\"{o}hl}},\ }\href@noop {}
  {\bibfield  {journal} {\bibinfo  {journal} {Nature}\ }\textbf {\bibinfo
  {volume} {464}},\ \bibinfo {pages} {388} (\bibinfo {year}
  {2010})}\BibitemShut {NoStop}%
\bibitem [{\citenamefont {Schmid}, \citenamefont {H\"arter},\ and\
  \citenamefont {Hecker~Denschlag}(2010)}]{Smi10}%
  \BibitemOpen
  \bibfield  {author} {\bibinfo {author} {\bibfnamefont {S.}~\bibnamefont
  {Schmid}}, \bibinfo {author} {\bibfnamefont {A.}~\bibnamefont {H\"arter}}, \
  and\ \bibinfo {author} {\bibfnamefont {J.}~\bibnamefont {Hecker~Denschlag}},\
  }\href {\doibase 10.1103/PhysRevLett.105.133202} {\bibfield  {journal}
  {\bibinfo  {journal} {Phys. Rev. Lett.}\ }\textbf {\bibinfo {volume} {105}},\
  \bibinfo {pages} {133202} (\bibinfo {year} {2010})}\BibitemShut {NoStop}%
\bibitem [{\citenamefont {Hall}\ \emph {et~al.}(2011)\citenamefont {Hall},
  \citenamefont {Aymar}, \citenamefont {Bouloufa-Maafa}, \citenamefont
  {Dulieu},\ and\ \citenamefont {Willitsch}}]{Hall2011}%
  \BibitemOpen
  \bibfield  {author} {\bibinfo {author} {\bibfnamefont {F.~H.~J.}\
  \bibnamefont {Hall}}, \bibinfo {author} {\bibfnamefont {M.}~\bibnamefont
  {Aymar}}, \bibinfo {author} {\bibfnamefont {N.}~\bibnamefont
  {Bouloufa-Maafa}}, \bibinfo {author} {\bibfnamefont {O.}~\bibnamefont
  {Dulieu}}, \ and\ \bibinfo {author} {\bibfnamefont {S.}~\bibnamefont
  {Willitsch}},\ }\href {\doibase 10.1103/PhysRevLett.107.243202} {\bibfield
  {journal} {\bibinfo  {journal} {Phys. Rev. Lett.}\ }\textbf {\bibinfo
  {volume} {107}},\ \bibinfo {pages} {243202} (\bibinfo {year}
  {2011})}\BibitemShut {NoStop}%
\bibitem [{\citenamefont {Rellergert}\ \emph {et~al.}(2011)\citenamefont
  {Rellergert}, \citenamefont {Sullivan}, \citenamefont {Kotochigova},
  \citenamefont {Petrov}, \citenamefont {Chen}, \citenamefont {Schowalter},\
  and\ \citenamefont {Hudson}}]{Hud2011}%
  \BibitemOpen
  \bibfield  {author} {\bibinfo {author} {\bibfnamefont {W.}~\bibnamefont
  {Rellergert}}, \bibinfo {author} {\bibfnamefont {S.}~\bibnamefont
  {Sullivan}}, \bibinfo {author} {\bibfnamefont {S.}~\bibnamefont
  {Kotochigova}}, \bibinfo {author} {\bibfnamefont {A.}~\bibnamefont {Petrov}},
  \bibinfo {author} {\bibfnamefont {K.}~\bibnamefont {Chen}}, \bibinfo {author}
  {\bibfnamefont {S.}~\bibnamefont {Schowalter}}, \ and\ \bibinfo {author}
  {\bibfnamefont {E.}~\bibnamefont {Hudson}},\ }\href@noop {} {\bibfield
  {journal} {\bibinfo  {journal} {Phys. Rev. Lett.}\ }\textbf {\bibinfo
  {volume} {107}},\ \bibinfo {pages} {243201} (\bibinfo {year}
  {2011})}\BibitemShut {NoStop}%
\bibitem [{\citenamefont {Ravi}\ \emph {et~al.}(2012)\citenamefont {Ravi},
  \citenamefont {Lee}, \citenamefont {Sharma}, \citenamefont {Werth},\ and\
  \citenamefont {Rangwala}}]{Ran2012}%
  \BibitemOpen
  \bibfield  {author} {\bibinfo {author} {\bibfnamefont {K.}~\bibnamefont
  {Ravi}}, \bibinfo {author} {\bibfnamefont {S.}~\bibnamefont {Lee}}, \bibinfo
  {author} {\bibfnamefont {A.}~\bibnamefont {Sharma}}, \bibinfo {author}
  {\bibfnamefont {G.}~\bibnamefont {Werth}}, \ and\ \bibinfo {author}
  {\bibfnamefont {S.~A.}\ \bibnamefont {Rangwala}},\ }\href {\doibase
  10.1038/ncomms2131} {\bibfield  {journal} {\bibinfo  {journal} {Nat.
  Commun.}\ }\textbf {\bibinfo {volume} {3}},\ \bibinfo {pages} {1126}
  (\bibinfo {year} {2012})}\BibitemShut {NoStop}%
\bibitem [{\citenamefont {Berkeland}\ \emph {et~al.}(1998)\citenamefont
  {Berkeland}, \citenamefont {Miller}, \citenamefont {Bergquist}, \citenamefont
  {Itano},\ and\ \citenamefont {Wineland}}]{Ber98}%
  \BibitemOpen
  \bibfield  {author} {\bibinfo {author} {\bibfnamefont {D.~J.}\ \bibnamefont
  {Berkeland}}, \bibinfo {author} {\bibfnamefont {J.~D.}\ \bibnamefont
  {Miller}}, \bibinfo {author} {\bibfnamefont {J.~C.}\ \bibnamefont
  {Bergquist}}, \bibinfo {author} {\bibfnamefont {W.~M.}\ \bibnamefont
  {Itano}}, \ and\ \bibinfo {author} {\bibfnamefont {D.~J.}\ \bibnamefont
  {Wineland}},\ }\href@noop {} {\bibfield  {journal} {\bibinfo  {journal} {J.
  Appl. Phys.}\ }\textbf {\bibinfo {volume} {83}},\ \bibinfo {pages} {10}
  (\bibinfo {year} {1998})}\BibitemShut {NoStop}%
\bibitem [{\citenamefont {Raab}\ \emph {et~al.}(2000)\citenamefont {Raab},
  \citenamefont {Eschner}, \citenamefont {Bolle}, \citenamefont {Oberst},
  \citenamefont {Schmidt-Kaler},\ and\ \citenamefont {Blatt}}]{Raa00}%
  \BibitemOpen
  \bibfield  {author} {\bibinfo {author} {\bibfnamefont {C.}~\bibnamefont
  {Raab}}, \bibinfo {author} {\bibfnamefont {J.}~\bibnamefont {Eschner}},
  \bibinfo {author} {\bibfnamefont {J.}~\bibnamefont {Bolle}}, \bibinfo
  {author} {\bibfnamefont {H.}~\bibnamefont {Oberst}}, \bibinfo {author}
  {\bibfnamefont {F.}~\bibnamefont {Schmidt-Kaler}}, \ and\ \bibinfo {author}
  {\bibfnamefont {R.}~\bibnamefont {Blatt}},\ }\href {\doibase
  10.1103/PhysRevLett.85.538} {\bibfield  {journal} {\bibinfo  {journal} {Phys.
  Rev. Lett.}\ }\textbf {\bibinfo {volume} {85}},\ \bibinfo {pages} {538}
  (\bibinfo {year} {2000})}\BibitemShut {NoStop}%
\bibitem [{\citenamefont {Chuah}\ \emph {et~al.}(2012)\citenamefont {Chuah},
  \citenamefont {Lewty}, \citenamefont {R.Cazan},\ and\ \citenamefont
  {Barrett}}]{Chu12}%
  \BibitemOpen
  \bibfield  {author} {\bibinfo {author} {\bibfnamefont {B.}~\bibnamefont
  {Chuah}}, \bibinfo {author} {\bibfnamefont {N.}~\bibnamefont {Lewty}},
  \bibinfo {author} {\bibnamefont {R.Cazan}}, \ and\ \bibinfo {author}
  {\bibfnamefont {M.}~\bibnamefont {Barrett}},\ }\href@noop {} {\bibfield
  {journal} {\bibinfo  {journal} {arXiv:1211.0101}\ } (\bibinfo {year}
  {2012})}\BibitemShut {NoStop}%
\bibitem [{\citenamefont {Pyka}\ \emph {et~al.}(2012)\citenamefont {Pyka},
  \citenamefont {Herschbach}, \citenamefont {Keller},\ and\ \citenamefont
  {Mehlst\"aubler}}]{Pyk12}%
  \BibitemOpen
  \bibfield  {author} {\bibinfo {author} {\bibfnamefont {K.}~\bibnamefont
  {Pyka}}, \bibinfo {author} {\bibfnamefont {N.}~\bibnamefont {Herschbach}},
  \bibinfo {author} {\bibfnamefont {J.}~\bibnamefont {Keller}}, \ and\ \bibinfo
  {author} {\bibfnamefont {T.}~\bibnamefont {Mehlst\"aubler}},\ }\href@noop {}
  {\bibfield  {journal} {\bibinfo  {journal} {arXiv:1206.5111}\ } (\bibinfo
  {year} {2012})}\BibitemShut {NoStop}%
\bibitem [{\citenamefont {Narayanan}\ \emph {et~al.}(2011)\citenamefont
  {Narayanan}, \citenamefont {Daniilidis}, \citenamefont {M\"oller},
  \citenamefont {Clark}, \citenamefont {Ziesel}, \citenamefont {Singer},
  \citenamefont {Schmidt-Kaler},\ and\ \citenamefont {H\"affner}}]{Nar11}%
  \BibitemOpen
  \bibfield  {author} {\bibinfo {author} {\bibfnamefont {S.}~\bibnamefont
  {Narayanan}}, \bibinfo {author} {\bibfnamefont {N.}~\bibnamefont
  {Daniilidis}}, \bibinfo {author} {\bibfnamefont {S.~A.}\ \bibnamefont
  {M\"oller}}, \bibinfo {author} {\bibfnamefont {R.}~\bibnamefont {Clark}},
  \bibinfo {author} {\bibfnamefont {F.}~\bibnamefont {Ziesel}}, \bibinfo
  {author} {\bibfnamefont {K.}~\bibnamefont {Singer}}, \bibinfo {author}
  {\bibfnamefont {F.}~\bibnamefont {Schmidt-Kaler}}, \ and\ \bibinfo {author}
  {\bibfnamefont {H.}~\bibnamefont {H\"affner}},\ }\href {\doibase
  10.1063/1.3665647} {\bibfield  {journal} {\bibinfo  {journal} {Journal of
  Applied Physics}\ }\textbf {\bibinfo {volume} {110}},\ \bibinfo {eid}
  {114909} (\bibinfo {year} {2011})}\BibitemShut {NoStop}%
\bibitem [{\citenamefont {Tanaka}\ \emph {et~al.}(2012)\citenamefont {Tanaka},
  \citenamefont {Masuda}, \citenamefont {Akimoto}, \citenamefont {Koda},
  \citenamefont {Ibaraki},\ and\ \citenamefont {Urabe}}]{Tan12}%
  \BibitemOpen
  \bibfield  {author} {\bibinfo {author} {\bibfnamefont {U.}~\bibnamefont
  {Tanaka}}, \bibinfo {author} {\bibfnamefont {K.}~\bibnamefont {Masuda}},
  \bibinfo {author} {\bibfnamefont {Y.}~\bibnamefont {Akimoto}}, \bibinfo
  {author} {\bibfnamefont {K.}~\bibnamefont {Koda}}, \bibinfo {author}
  {\bibfnamefont {Y.}~\bibnamefont {Ibaraki}}, \ and\ \bibinfo {author}
  {\bibfnamefont {S.}~\bibnamefont {Urabe}},\ }\href {\doibase
  10.1007/s00340-011-4762-2} {\bibfield  {journal} {\bibinfo  {journal}
  {Applied Physics B}\ }\textbf {\bibinfo {volume} {107}},\ \bibinfo {pages}
  {907} (\bibinfo {year} {2012})}\BibitemShut {NoStop}%
\bibitem [{\citenamefont {Zipkes}\ \emph {et~al.}(2011)\citenamefont {Zipkes},
  \citenamefont {Ratschbacher}, \citenamefont {Sias},\ and\ \citenamefont
  {K\"{o}hl}}]{Zip11}%
  \BibitemOpen
  \bibfield  {author} {\bibinfo {author} {\bibfnamefont {C.}~\bibnamefont
  {Zipkes}}, \bibinfo {author} {\bibfnamefont {L.}~\bibnamefont
  {Ratschbacher}}, \bibinfo {author} {\bibfnamefont {C.}~\bibnamefont {Sias}},
  \ and\ \bibinfo {author} {\bibfnamefont {M.}~\bibnamefont {K\"{o}hl}},\
  }\href@noop {} {\bibfield  {journal} {\bibinfo  {journal} {New. J. Phys.}\
  }\textbf {\bibinfo {volume} {13}},\ \bibinfo {pages} {053020} (\bibinfo
  {year} {2011})}\BibitemShut {NoStop}%
\bibitem [{\citenamefont {Schmid}\ \emph {et~al.}(2012)\citenamefont {Schmid},
  \citenamefont {H\"arter}, \citenamefont {Frisch}, \citenamefont {Hoinka},\
  and\ \citenamefont {Hecker~Denschlag}}]{Smi12}%
  \BibitemOpen
  \bibfield  {author} {\bibinfo {author} {\bibfnamefont {S.}~\bibnamefont
  {Schmid}}, \bibinfo {author} {\bibfnamefont {A.}~\bibnamefont {H\"arter}},
  \bibinfo {author} {\bibfnamefont {A.}~\bibnamefont {Frisch}}, \bibinfo
  {author} {\bibfnamefont {S.}~\bibnamefont {Hoinka}}, \ and\ \bibinfo {author}
  {\bibfnamefont {J.}~\bibnamefont {Hecker~Denschlag}},\ }\href@noop {}
  {\bibfield  {journal} {\bibinfo  {journal} {Rev. Sci. Instrum.}\ }\textbf
  {\bibinfo {volume} {83}},\ \bibinfo {pages} {053108} (\bibinfo {year}
  {2012})}\BibitemShut {NoStop}%
\bibitem [{\citenamefont {H\"arter}\ \emph {et~al.}(2012)\citenamefont
  {H\"arter}, \citenamefont {Kr\"ukow}, \citenamefont {Brunner}, \citenamefont
  {Schnitzler}, \citenamefont {Schmid},\ and\ \citenamefont
  {Hecker~Denschlag}}]{Har12}%
  \BibitemOpen
  \bibfield  {author} {\bibinfo {author} {\bibfnamefont {A.}~\bibnamefont
  {H\"arter}}, \bibinfo {author} {\bibfnamefont {A.}~\bibnamefont {Kr\"ukow}},
  \bibinfo {author} {\bibfnamefont {A.}~\bibnamefont {Brunner}}, \bibinfo
  {author} {\bibfnamefont {W.}~\bibnamefont {Schnitzler}}, \bibinfo {author}
  {\bibfnamefont {S.}~\bibnamefont {Schmid}}, \ and\ \bibinfo {author}
  {\bibfnamefont {J.}~\bibnamefont {Hecker~Denschlag}},\ }\href@noop {}
  {\bibfield  {journal} {\bibinfo  {journal} {Phys. Rev. Lett.}\ }\textbf
  {\bibinfo {volume} {109}},\ \bibinfo {pages} {123201} (\bibinfo {year}
  {2012})}\BibitemShut {NoStop}%
\bibitem [{Note1()}]{Note1}%
  \BibitemOpen
  \bibinfo {note} {To calculate $E_\protect \text {z}$ we use equation$\protect
  \tmspace +\thinmuskip {.1667em}$\ref {Eemm} and replace $q_r$ and $r$ by
  $q_z$ and $z$. In our trap $q_z\approx 0.04\cdot q_r\approx
  0.01$.}\BibitemShut {Stop}%
\bibitem [{\citenamefont {Chou}\ \emph {et~al.}(2010)\citenamefont {Chou},
  \citenamefont {Hume}, \citenamefont {Koelemeij}, \citenamefont {Wineland},\
  and\ \citenamefont {Rosenband}}]{Cho10}%
  \BibitemOpen
  \bibfield  {author} {\bibinfo {author} {\bibfnamefont {C.~W.}\ \bibnamefont
  {Chou}}, \bibinfo {author} {\bibfnamefont {D.~B.}\ \bibnamefont {Hume}},
  \bibinfo {author} {\bibfnamefont {J.~C.~J.}\ \bibnamefont {Koelemeij}},
  \bibinfo {author} {\bibfnamefont {D.~J.}\ \bibnamefont {Wineland}}, \ and\
  \bibinfo {author} {\bibfnamefont {T.}~\bibnamefont {Rosenband}},\ }\href
  {\doibase 10.1103/PhysRevLett.104.070802} {\bibfield  {journal} {\bibinfo
  {journal} {Phys. Rev. Lett.}\ }\textbf {\bibinfo {volume} {104}},\ \bibinfo
  {pages} {070802} (\bibinfo {year} {2010})}\BibitemShut {NoStop}%
\bibitem [{\citenamefont {Akerman}(2012)}]{Ake12}%
  \BibitemOpen
  \bibfield  {author} {\bibinfo {author} {\bibfnamefont {N.}~\bibnamefont
  {Akerman}},\ }\emph {\bibinfo {title} {{Trapped ions and free photons}}},\
  \href@noop {} {Ph.D. thesis},\ \bibinfo  {school} {Weizmann Institute of
  Science Rehovot} (\bibinfo {year} {2012})\BibitemShut {NoStop}%
\bibitem [{\citenamefont {Chwalla}(2009)}]{Chw09}%
  \BibitemOpen
  \bibfield  {author} {\bibinfo {author} {\bibfnamefont {M.}~\bibnamefont
  {Chwalla}},\ }\emph {\bibinfo {title} {{Precision spectroscopy with
  $^{40}$Ca$^+$ ions in a Paul trap}}},\ \href@noop {} {Ph.D. thesis},\
  \bibinfo  {school} {Universit\"{a}t Innsbruck} (\bibinfo {year}
  {2009})\BibitemShut {NoStop}%
\bibitem [{\citenamefont {Hemmerling}(2011)}]{Hem11}%
  \BibitemOpen
  \bibfield  {author} {\bibinfo {author} {\bibfnamefont {B.}~\bibnamefont
  {Hemmerling}},\ }\emph {\bibinfo {title} {{Towards Direct Frequency Comb
  Spectroscopy Using Quantum Logic}}},\ \href@noop {} {Ph.D. thesis},\ \bibinfo
   {school} {Gottfried Wilhelm Leibniz Universit\"at Hannover} (\bibinfo {year}
  {2011})\BibitemShut {NoStop}%
\end{thebibliography}%

\end{document}